\documentclass[12pt,letterpaper]{article}
\usepackage{amsmath,amssymb,calc, amsthm}
\usepackage{graphicx}
\usepackage[nosort]{cite}
\usepackage{epsfig,psfrag}

\makeatletter
\renewcommand\section{\@startsection {section}{1}{\z@}%
                                   {-3.5ex \@plus -1ex \@minus -.2ex}%nn
                                   {2.3ex \@plus.2ex}%
                                   {\normalfont\large\bfseries}}
\renewcommand\subsection{\@startsection{subsection}{2}{\z@}%
                                     {-3.25ex\@plus -1ex \@minus -.2ex}%
                                     {1.5ex \@plus .2ex}%
                                     {\normalfont\bfseries}}
\makeatother

\theoremstyle{plain}

\theoremstyle{definition}

\let\non\nonumber

\let\w=\wedge

\def\one{^{(1)}}

\newcommand{\bea}{\begin{eqnarray}}
\newcommand{\eea}{\end{eqnarray}}
\newcommand{\be}{\begin{equation}}
\newcommand{\ee}{\end{equation}}
\newcommand{\bma}{\begin{pmatrix}}
\newcommand{\ema}{\end{pmatrix}}
\newcommand{\cc}{{\rm c.c.}}

\newcommand{\Z}{{\mathbb Z}}
\newcommand{\R}{{\mathbb R}}

\newcommand{\PP}{{\mathbb P}}

\newcommand{\cM}{{\cal M}}
\newcommand{\cR}{{\cal R}}

\newcommand{\ve}{\varepsilon}

\newcommand{\C}[1]{$(\ref{#1})$}

\def\IZ{\relax\ifmmode\mathchoice
{\hbox{\cmss Z\kern-.4em Z}}{\hbox{\cmss Z\kern-.4em Z}}
{\lower.9pt\hbox{\cmsss Z\kern-.4em Z}} {\lower1.2pt\hbox{\cmsss
Z\kern-.4em Z}}\else{\cmss Z\kern-.4em Z}\fi}
\def\IR{\relax{\rm I\kern-.18em R}}

\def\one{{\hbox{ 1\kern-.8mm l}}}

\def\tr{{\rm tr\,}}
\def\Tr{{\rm Tr\,}}

\newlength{\bredde}
\def\slash#1{\settowidth{\bredde}{$#1$}\ifmmode\,\raisebox{.15ex}{/}
\hspace*{-\bredde} #1\else$\,\raisebox{.15ex}{/}\hspace*{-\bredde}
#1$\fi}

\newsavebox{\zzzbar}
\sbox{\zzzbar}
  {\setlength{\unitlength}{0.9em}
  \begin{picture}(0.6,0.7)
  \thinlines
  \put(0,0){\line(1,0){0.6}}
  \put(0,0.75){\line(1,0){0.575}}
  \multiput(0,0)(0.0125,0.025){30}{\rule{0.3pt}{0.3pt}}
  \multiput(0.2,0)(0.0125,0.025){30}{\rule{0.3pt}{0.3pt}}
  \put(0,0.75){\line(0,-1){0.15}}
  \put(0.015,0.75){\line(0,-1){0.1}}
  \put(0.03,0.75){\line(0,-1){0.075}}
  \put(0.045,0.75){\line(0,-1){0.05}}
  \put(0.05,0.75){\line(0,-1){0.025}}
  \put(0.6,0){\line(0,1){0.15}}
  \put(0.585,0){\line(0,1){0.1}}
  \put(0.57,0){\line(0,1){0.075}}
  \put(0.555,0){\line(0,1){0.05}}
  \put(0.55,0){\line(0,1){0.025}}
  \end{picture}}

\newcommand{\ena}{\end{eqnarray}}
\newcommand{\beqa}{\begin{eqnarray}}
\newcommand{\eeqa}{\end{eqnarray}}

\newcommand{\half}{\frac{1}{2}}

\newfont{\goth}{ygoth.tfm scaled 1200}                   % gothic font (usual)
 \numberwithin{equation}{section}

\def\1{{(1)}}
\def\2{{(2)}}
\def\3{{(3)}}

\def\1{{\bf 1}}

\def\M{{\mathcal M}}

\def\1{{\bf 1}}
\def\3{{\bf 3}}
\def\7{{\bf 7}}
\def\2{{\bf 2}}
\def\8{{\bf 8}}

\newcommand{\nc}{\newcommand}
\nc{\cL}{{\cal L}}
\nc{\betabar}{\ol{\beta}}
\nc{\ol}{\overline}

\def\om{\omega}

\begin{document}
\begin{titlepage}

\begin{center}

%\today 
{October 4, 2012}
\hfill         \phantom{xxx}  EFI-12-13, \, DAMTP-2012-54, \, DMUS-MP-12/06

\vskip 2 cm {\Large \bf M-theory and Type IIA Flux Compactifications}
%\vskip 0.3 cm {\Large \bf K\"ahler Potentials}\non\\
\vskip 1.25 cm {\bf  Jock McOrist$^{a, b}$\footnote{j.mcorist@damtp.cam.ac.uk}\ and Savdeep Sethi$^{c}$\footnote{sethi@uchicago.edu}}\non\\
\vskip 0.5cm  $^{a}${\it Department of Applied Mathematics and Theoretical Physics, Centre for Mathematical Sciences, Wilberforce Road, Cambridge, CB3 0WA, UK} \non\\
\vskip 0.2 cm
$^{b}${\it Department of Mathematics, University of Surrey,
Guildford GU2 7XH, UK} \non\\
\vskip 0.2 cm
$^{c}${\it Enrico Fermi Institute, University of Chicago, Chicago, IL 60637, USA}
\end{center}
\vskip 1.5 cm

\begin{abstract}
\baselineskip=18pt

We consider compactifications of M-theory and type IIA string theory to four dimensions. %If space-time is Minkowski,
For Minkowski space-time, a supergravity no-go theorem forbids flux supported in the internal space. We show how to evade this no-go theorem by exhibiting new sources of brane charge: in string theory, the basic physical phenomenon is the generation of new brane charges from D-branes in transverse fluxes. In M-theory, there is a new source of M5-brane charge from novel higher derivative couplings that involve fluxes as well as curvatures. We present some explicit orientifold examples with both ${\cal N}=1$ and ${\cal N}=2$ space-time supersymmetry. Finally, we explain the status of massive type IIA flux compactifications.

\end{abstract}

\end{titlepage}

%\tableofcontents

\section{Introduction}
\label{intro}

Fluxes are a  crucial ingredient in the construction of four-dimensional string vacua. At the level of supergravity, a no-go theorem by Gibbons forbids static flux compactifications to Minkowski or de Sitter space~\cite{Gibbons:1984kp}.   This constraint, which follows from the Einstein equations, is of Gauss law type and holds regardless of whether supersymmetry is preserved or broken. Particularly for supersymmetric backgrounds, this constraint can be related to charge conservation for an appropriate brane charge. The known supersymmetric flux solutions are of the type where supergravity fluxes and explicit brane sources are  balanced against some higher derivative source of brane charge.

The best understood case is the heterotic string for which the relevant conserved charge is NS5-brane charge. There exists a gravitational correction to the Bianchi identity for the Neveu-Schwarz (NS) $3$-form ${\cal H}_3$,
\be \label{hetbianchi}
d {\cal H}_3 = { \alpha' \over 4} \left( \Tr (R \wedge R) - \Tr (F \wedge F) \right),
\ee
which is a four derivative interaction that induces a background NS5-brane charge tadpole. There are also associated four derivative $R^2$-type couplings that permit one to evade the supergravity constraints prohibiting a background ${\cal H}_3$. The price that is paid in the heterotic string is a potentially small volume string compactification.

%This source for the Bianchi identity allows one to turn on ${\cal H}_3$ and obtain compact non-K\"ahler heterotic solutions~\cite{Dasgupta:1999ss}.

Flux models with large volume limits were originally found in compactifications of M-theory to three dimensions~\cite{Becker:1996gj}, and type IIB or F-theory compactifications to four dimensions~\cite{Sethi:1996es, Dasgupta:1999ss}. In the former case, there is an M2-brane charge tadpole while in the latter, a D3-brane charge tadpole. The charge tadpole has the same origin in both cases: namely higher derivative couplings in the M-theory effective action. In type IIB string theory, these higher derivative interactions are supported on branes and orientifold planes. In addition to the charge terms, there are again associated higher derivative gravitational couplings like $R^4$-type terms in three dimensions, or $R^2$-type terms in four dimensions, which modify the two derivative supergravity constraints.

There are a few points that are worth explaining. F-theory backgrounds are typically not solutions of type IIB supergravity. They include $(p,q)$ $7$-brane sources~\cite{Vafa:1996xn}. Our belief in the existence of such backgrounds is not from any perturbative string argument, but because these solutions are found as limits of smooth elliptic M-theory geometries. The same statement applies to flux vacua. In compactifications of M-theory to $3$-dimensions, the higher derivative couplings allow smooth flux backgrounds. Suitable elliptic models have a type IIB limit, and this provides the most compelling evidence for the existence of F-theory flux vacua.
We should note that the orientifold limit of F-theory vacua is the only limit in which the string coupling is constant over the compactification space~\cite{Sen:1996vd}. The M-theory metric  corresponding to this orientifold limit has only orbifold singularities.  In this limit, the charge and tension of each $O7^-$-plane is canceled point-wise by D-branes. There are no naked exotic sources like $O$-planes. The notation $Op^-$ refers to the orientifold plane that gives $SO(2N)$ gauge symmetry when collided with D$p$-branes. This orientifold plane has both negative charge and negative tension. In principle, these backgrounds can be understood in string perturbation theory, although there is no tunable string coupling since the dilaton is determined by the fluxes~\cite{Dasgupta:1999ss}. Despite much effort, we currently lack a useful definition of string theory in these backgrounds.

%Orientifold limits of F-theory are of this type. Both the string coupling and warp factor are  well-behaved in this class of compactifications. Metric singularities are typically no worse than orbifold singularities, and the models can be reasonably studied in supergravity together with higher derivative couplings supported on the D-branes and orientifolds. These higher derivative couplings can produce tadpoles for lower brane charges; in type IIB, couplings on the D$7$-branes and $O7$-planes produce a D3-brane tadpole. Given the difficulty in understanding perturbative string theory  in the presence of fluxes, the most compelling reason to believe in the existence of type IIB flux vacua is because they are limits of smooth M-theory flux vacua.

We can now describe the status of M-theory and type IIA compactifications to four dimensions with flux. First note that there is no problem turning on $F_2$-flux in type IIA. Unlike higher form flux, $F_2$-flux descends from a pure metric configuration in M-theory, with the choice of $F_2$-flux determined by the topology of the circle bundle on which we reduce from M-theory to string theory. As long as the seven-dimensional metric solves the supergravity equations of motion, the corresponding IIA flux is certainly permitted. The really interesting question involves $4$-form flux for which there is a basic mystery: how in these theories can we avoid the supergravity no-go theorems that prohibit fluxes?

One way is to consider spaces with boundaries supporting extra degrees of freedom. For example type IIA on $S^1/\Z_2$ with D8-branes and $O8$-planes certainly permits $4$-form flux. Orientifolds of this type are often dual to type I backgrounds. One meets an example of this type in~\cite{Dasgupta:1999ss}\ on route to a smooth torsional type I/heterotic background. Another example is the strong coupling limit of the $E_8\times E_8$ heterotic string on a $CY_3$, given by M-theory on $S^1/\Z_2 \times CY_3$. This background certainly includes $4$-form flux described in~\cite{Witten:1996mz}. Each of these cases basically involves a strong coupling analogue of a heterotic or type I mechanism permitting flux. In this work, we are primarily interested in explaining the ingredients needed for M-theory and type IIA backgrounds with $4$-form flux, but without boundaries. For this reason, we will exclude type IIA backgrounds with $O8$-planes.

There is a another possibility. The orientifold limit of F-theory involves a point-wise cancelation of the orientifold charge and tension against D-branes. The required D3-brane charge tadpole is produced from gravitational couplings supported on the D7-branes and $O7$-planes. One could also consider backgrounds with naked orientifold planes like $O3$ and $O5$-planes in type IIB or $O4$ and $O6$-planes in type IIA. The orientifold charge and tension can sometimes be balanced against supergravity fluxes, but the cancelation is not point-wise. These vacua always contain singular sources at the level of supergravity. The nature of these singular sources depends on the particular $O$-plane. Near each plane, some form of strong coupling physics must resolve the singularity. String perturbation theory automatically provides a definition of this background when such a description is available. Otherwise, we must rely on some space-time understanding of the physics which resolves the singularity. Such an understanding exists via M-theory for $O6^-$-planes to be discussed in section~\ref{o6enough}. An analogous smoothing of the metric does not happen for type IIA $Op^-$-planes with $p<6$. For example, $O4^-$ lifts to the M-theory orientifold $\R^5/\Z_2 \times S^1$~\cite{Dasgupta:1995zm, Witten:1995em, Hori:1998iv}, while $O2^-$ lifts to the M-theory orbifold $\R^8/\Z_2$~\cite{Sethi:1998zk, Berkooz:1998sn}.\footnote{An orientifold of M-theory is a quotient action that includes inversion of the $3$-form potential. } Both configurations are singular M-theory backgrounds.

We certainly expect backgrounds to exist with naked $O3^-$ and $O4^-$-planes. Flux models with $O3$-planes were considered in~\cite{Dasgupta:1999ss, Frey:2002hf, Kachru:2002he}. However, they are on a somewhat different footing from vacua involving $O6^-$ and $O7^-$-planes because one needs to understand something about singularities in M-theory. We really seek a picture for type IIA and M-theory flux vacua similar to the situation for F-theory vacua. This will require new sources of brane charge on smooth backgrounds so $O4^-$-planes of type IIA cannot play any essential role in the basic mechanism.

\subsection{Are $O6$-planes enough?} \label{o6enough}

At first sight, type IIA and type IIB look very similar. The only new ingredient needed in type IIB to construct flux backgrounds are $O7$-planes or their F-theory generalizations. These are the sources which induce a D3-brane charge tadpole in a pure metric background. In analogy, one might imagine that considering an orientifold of IIA that produces $O6$-planes would be sufficient to permit fluxes.\footnote{This erroneous claim can even be found in review papers on this topic.} %Note that $Op$-planes for $p<6$ with negative tension are typically singular configurations both at weak and strong coupling.
 It is not hard to see that this cannot be the case for any of the orientifold six-planes that are believed to exist. In string theory, $O6$-planes appear in potentially four flavors, labeled $O6^-, O6^{-'}, O6^+, O6^{+'}$ in~\cite{Boer:2002yq}. The $O6^-$ and $O6^+$-planes are found in conventional type IIA perturbative string theory. In the presence of coincident D6-branes, these planes respectively give rise to $SO(2N)$ or $Sp(N)$ gauge symmetries.

However in both cases, we know something about the strong coupling lift to M-theory.  In the case of $O6^-$, the lift is the smooth Atiyah-Hitchin manifold~\cite{Seiberg:1996bs,Seiberg:1996nz} while the $O6^+$ lifts to a kind of frozen $D_4$ singularity~\cite{Landsteiner:1997ei, Witten:1997bs}. As a reminder,  we note that the lift of a single D6-brane is also smooth geometry, namely a Taub-NUT space. That description will be very useful for us later. Although the singularity is frozen in the $O6^+$ case, it is not morally different from a conventional $ALE$ space for the purpose of providing a violation of energy conditions; this can be seen from the dual heterotic description which is quite conventional and from the fact that the tension is positive in string theory. Therefore, both cases lift to seven-dimensional M-theory backgrounds without sufficiently exotic ingredients to violate the constraints prohibiting flux.  In the case of $O6^-$ and D6-branes, there are no exotic ingredients whatsoever.

This leaves two orientifolds to consider which are more peculiar: $(O6^{-'}, O6^{+'})$. These orientifolds do not exist in conventional perturbative string theory. If they exist at all, it must be in massive type IIA with odd cosmological constant~\cite{Hyakutake:2000mr}. The only orientifolds compatible with even cosmological constant are $(O6^{-}, O6^{+})$. This follows from T-duality and an essentially topological argument. However, it is quite unclear how such orientifolds are to be defined. Orientifolds are intrinsically stringy objects and no string theory description currently exists for massive type IIA supergravity. This issue is further discussed in section~\ref{massive}.

For most of our discussion, it does not matter whether orientifold planes can be defined in massive type IIA. Aside from section~\ref{massive}, we will only discuss conventional type IIA string theory.
%Without some additional ingredient, $O6$-planes in massive IIA are no more able to evade the supergravity no-go theorem than $O6$-planes in conventional type IIA string theory.
As we will show, the necessary new ingredient in type IIA and M-theory is flux itself. A combination of flux and metric can generate a charge tadpole that permits evasion of the supergravity no-go theorem. In this respect, M-theory and type IIA are starkly different from type IIB, heterotic and type I string theory. It is reasonable to expect a similar picture for massive IIA.

%One might also imagine type IIA compactifications which include $O4$-planes, where the orientifold charge is canceled by a combination of fluxes. These compactifications are singular at the level of supergravity. The M-theory description of $O4^-$, with negative tension, is the orbifold $R^5/\Z_2$ which is also singular.

\subsection{The basic idea and relation to past work}

We will mainly concern ourselves with compactifications to four-dimensional Minkowski or $AdS$ space-time. In section~\ref{nogo}, we describe the basic supergravity constraints on such compactifications. These constraints do not require supersymmetry. Compactifications to $AdS_4$  of Freund-Rubin type are certainly possible for both M-theory and type IIA~\cite{Freund:1980xh}. In those cases, the scale of the $AdS_4$ space is typically of order the compactification scale so one should view the supergravity background as either ten or eleven-dimensional.

Our goal is to find four-dimensional compactifications with either Minkowski space-time, or a large separation between the $AdS_4$ scale and the Kaluza-Klein scale. There is no precise no-go theorem for $AdS_4$ with a large separation of scales, but the intuition is that one should expect to encounter the same obstructions present for Minkowski space-time when the scale separation can be made parametrically large.

Let us focus on M-theory for the moment. To find Minkowski solutions, we need a charge tadpole. The only relevant charge is M5-brane charge so there must be new contributions to the Bianchi identity determining M5-brane charge. To find these contributions, we will start with the source of M2-brane charge that makes possible $3$-dimensional M-theory flux compactifications. This source is the gravitational $8$ derivative coupling in the M-theory effective action,
\be\label{m2source}
\int C_3 \wedge X_8.
\ee
The basic idea goes as follows: while string duality symmetries can mix large and small volumes, it is reasonable to expect the total charge violation to be robust under duality.  This is essentially because tadpoles (like anomalies) are one-loop effects. With this in mind, we dualize~\C{m2source}\ into an $8$ derivative M-theory coupling applicable to seven-dimensional spaces sourcing M5-brane charge:
\be
\int C_6 \wedge X_5.
\ee
This dualization is carried out in section~\ref{dualizing}.  A related set of calculations can be found in~\cite{Liu:2010gz}.\footnote{We would like to thank Ruben Minasian for bringing this work to our attention.} The resulting $X_5$ is not a purely gravitational coupling, but is constructed from metrics and fluxes. It is quite strange partly because it is an odd-dimensional class. It is this coupling that generates the desired tadpole in M-theory.

Along the way, we will find similar couplings on D-branes, and specifically, D6-branes. There are many directions to explore. It is going to be very interesting to classify all such couplings on branes which can generate physical charge. All of these couplings are crying out for a natural geometric home, which is likely to involve a better understanding of Dirac quantization and anomaly cancelation in the presence of flux.

In section~\ref{compact}, we use a duality chain to construct some explicit M-theory orientifold examples of Minkowski flux vacua with both ${\cal N}=1$ and ${\cal N}=2$ supersymmetry. Constructing smooth examples that are not orientifolds remains an outstanding open question. It appears this should be possible, but the question is quite non-trivial. There is still quite some work needed to state a completely general $7$-dimensional M-theory metric and flux ansatz, analogous to the F-theory case of conformal $CY_4$ with a choice of $4$-form flux, which will satisfy both the supersymmetry constraints (or at least the equations of motion) and the tadpole conditions. However, this now appears to be a tractable question.

There is also an interesting question about whether these new couplings might allow de Sitter solutions, which are ruled out at the level of supergravity. The same question can actually be asked of type IIB flux vacua and even M-theory $3$-dimensional flux vacua. We suspect this will not be possible, but an analysis analogous to the one performed for the heterotic string in~\cite{Green:2011cn, Gautason:2012tb}\ is needed.

Finally, in section~\ref{massive}\ we turn to the remaining corner of the string landscape which is massive type IIA. This is the only corner in which there is a proposed mechanism for compact flux vacua  different in nature from the mechanisms in heterotic and type IIB string theory, and quite different from our proposal for M-theory and conventional type IIA. The proposal by~\cite{DeWolfe:2005uu}\ is for $AdS_4$ vacua based on $CY_3$ compactifications with very striking features. Namely, a parametrically large separation between the $AdS_4$ scale and the Kaluza-Klein scale along with a large internal volume and a small string coupling. The parameter is the amount of $4$-form flux in the internal space. The only new ingredient beyond massive IIA supergravity are $O6$-planes. We describe some of the problems with this proposal and show that the proposed backgrounds do not provide approximate solutions of massive IIA. This leaves massive IIA as an open area for which an understanding of compact flux vacua is still lacking. With a proper understanding of tadpole constraints, we suspect the picture for massive IIA will more closely resemble other corners of string theory.

\section{Supergravity Constraints}
\label{nogo}

Let us begin with eleven-dimensional supergravity which is the cleanest case. The theory contains a metric $g$ and a $3$-form potential $C$ with field strength $G$. In later sections when discussing string theory, we will use a subscript $C_p$ to denote a $p$-form potential. The bosonic terms in the supergravity action take the form
\be
S = \frac{1}{2\kappa^2} \int d^{11}x \sqrt{-g} \left(  \cR - \frac{1}{2} |G|^2\right)  - \frac{1}{2\kappa^2} \int \frac{1}{6} C \w G \w G ,\label{eqn:action}
\ee
where $\cR_{MN}$ is the Ricci tensor and $\cR = g^{MN} \cR_{MN}$ is the Ricci scalar. From now on, we will set $\kappa=1$. The equations of motion that follow from \C{eqn:action} are:
\bea
\cR_{MN}    &=& \frac{1}{12} \left( G_{MPQR} G_{N}^{~~PQR} -2\,g_{MN} |G|^2\right).\label{eqn:einstein}\\
d\star G &=& -\half G \w G,\label{f4eom}
\eea
where the norm of a rank $p$ tensor is defined by
\be\non
|T|^2 = {1\over p!} g^{M_1N_1} \cdots g^{M_p N_p} T_{M_1\ldots M_p} T_{N_1 \ldots N_p}.
\ee In the absence of explicit M5-brane sources, the field strength also satisfies the Bianchi identity $dG = 0$.

We  assume our space-time is a warped product of a maximally symmetric space-time with a compact $7$-manifold, $\cM_4 \times_w \cM_7$, and
correspondingly consider a metric of the form
\be ds^2 = g_{MN} dx^M dx^N = e^{2w(y)} \left( \hat{g}_{\mu\nu}(x) dx^\mu dx^\nu + \hat{g}_{mn}(y) dy^m dy^n \right), \label{eqn:metric_1} \ee
with ${\hat g}_{\mu\nu}$ the unwarped metric of $\cM_4$.
Any $4$-form flux must be compatible with this ansatz and must therefore take the form \bea G_{mnpq}, \quad
G_{\mu\nu\rho\tau} = f \sqrt{-g^{(4)}} \, \ve_{\mu\nu\rho\tau}.\label{eqn:four-form}
\eea The space-time flux is proportional to the volume form of $\cM_4$. In principle, the Freund-Rubin parameter $f$ can depend on the
internal coordinates. However, if we insist on no M5-brane sources so that
\be
dG =0,
\ee
then $f=f_0 e^{-4w}$ with $f_0$ constant.

The flux equation of motion then provides an interesting constraint. The $G\wedge G$ interaction is only non-vanishing for the combination $G^{(4)} \wedge G^{\rm int}$ where $G^{(4)}$ refers to the space-time flux and $G^{\rm int}$ to the internal flux. This gives a relation:
\be
d\left( e^{-3w}\, {\hat \ast}_7 G^{\rm int} \right) = - f_0 G^{\rm int}.
\ee
For $(w=0, f_0=0)$, this is standard harmonicity for the flux $G^{\rm int}$.
Tracing the Einstein equations~\C{eqn:einstein}\ relates the total scalar curvature to the flux,
\be
\cR  = {1\over 6} |G|^2, \qquad |G|^2 = |G^{(4)}|^2 + |G^{\rm int}|^2, \qquad |G^{(4)}|^2 = -f^2.
\ee
Separately tracing over the internal and space-time metrics provides relations on the warped four and seven-dimensional scalar curvatures, $\cR^{(4)}$ and $\cR^{(7)}$, respectively:
\be\label{warpedcurvatures}
\cR^{(4)} = - {4\over 3} f^2 - {2\over 3} |G^{\rm int}|^2, \qquad \cR^{(7)} = {5\over 6} |G^{\rm int}|^2 + {7\over 6}  f^2.
\ee

To understand the implication for the unwarped space-time metric, it is useful to rewrite the Ricci curvature for a $D$-dimensional space in terms of hatted unwarped quantities by a conformal transformation:
\be
\cR_{MN}= \hat{\cR}_{MN} - \hat{g}_{MN} \hat{\nabla}^2 w +(D-2)\left( \hat{\nabla}_M w \hat{\nabla}_N w -  \hat{\nabla}_M  \hat{\nabla}_N w - \hat{g}_{MN} |  \hat{\nabla}_P w|^2 \right).
\ee
For a maximally symmetric space-time like $AdS_4$ or Minkowski space-time, we require ${\hat \cR}_{\mu\nu} =  {\hat g}_{\mu\nu} \Lambda$ with $\Lambda$ constant. First let us study the scalar curvature rather than the Ricci tensor to find a global constraint on $\Lambda$. Using $D=11$ and tracing over the four-dimensional space-time indices gives,
\bea
\cR^{(4)} &=& e^{-2w}\left(  \hat{\cR}^{(4)}  - 4 \hat{\nabla}^2 w -36|  \hat{\nabla} w|^2 \right) \cr
&=& 4e^{-2w}\left(  \Lambda -{1\over 9} e^{-9w}  \hat{\nabla}^2 e^{9w} \right) \cr
&\leq& 0,
\eea
with $\Lambda$ the four-dimensional cosmological constant, and the last inequality following from \C{warpedcurvatures}. From the expression,
\be\label{lambda}
\Lambda = {1\over 4} e^{2w} \cR^{(4)}  + {1\over 9} e^{-9w}\hat{\nabla}^2 e^{9w} \quad \Rightarrow \quad \Lambda = {\int_{\cM_7}  e^{11w} \,  \cR^{(4)} \over    4\int_{\cM_7}  e^{9w} },
\ee
it is easy to see that $\Lambda=0$ implies $\cR^{(4)}=0$.  The integration measure in~\C{lambda}\ is with respect to the unwarped metric. In turn, $\cR^{(4)}=0$ requires all $G$-flux to vanish from~\C{warpedcurvatures}. This is the basic no-go result on Minkowski or de Sitter compactifications with flux. In general, $\Lambda \leq 0$.

The strongest constraint on internal flux comes from the space-time components of the Einstein equations~\C{eqn:einstein}.  Rewriting~\C{eqn:einstein}\ in terms of unwarped quantities gives
\be
{\hat \cR}_{\mu\nu} =   {\hat g}_{\mu\nu} \Lambda= {\hat g}_{\mu\nu} \left( {\hat\nabla}^2 w+ 9  | {\hat\nabla} w|^2 - {71\over 144} f_0^2 e^{-6w} - {1\over 144} e^{2w} |G^{\rm int}|^2 \right).
\ee
This is sharply restrictive since the right hand side must be point-wise independent of the $y$ coordinates, which requires a precise cancelation between the warp factor and the flux terms. For example, Freund-Rubin solutions correspond to constant warp factor $w$ and $G^{\rm int}=0$. There are solutions with $|G^{\rm int}|^2$ constant and non-vanishing with constant $w$, like the case studied in~\cite{Englert:1982vs}. There are also known solutions with a non-constant warp factor corresponding to compactification on a deformed $S^7$; for a review, see~\cite{Duff:1986hr}.

For a generic solution, space-time is $AdS_4$ with $\Lambda$ of order the curvature scale $\cR^{(7)}$. Flux solutions with parametrically small $\Lambda$ compared with $\cR^{(7)}$ will run into problems analogous to those encountered when trying to construct a pure Minkowski flux background. The solutions  are then either Minkowski with no flux, or $AdS_4$ Freund-Rubin $11$-dimensional solutions with a cosmological constant of order the Kaluza-Klein scale. This is a quite robust  picture which does not  require a detailed study of the Einstein equations. Note that conventional type IIA supergravity is a special case of this discussion. To evade this no-go result, we will need ingredients beyond supergravity.

\section{New Couplings from Duality}
\label{dualizing}
\subsection{Known M-theory couplings}

To evade the constraints of section~\ref{nogo}, we need sources that can
act like negative tension objects, and which can modify the flux
equations of motion. In principle, the higher derivative corrections to
supergravity found in M-theory can provide these sources.

The leading corrections to the supergravity action~\C{eqn:action}\ in a momentum expansion
are terms with $8$ derivatives, which are down by $\ell_p^6$
from the two derivative terms. Unfortunately, the complete action at
this order is currently unknown. However, specific couplings are known which
take the form
\be \label{oldcouplings} S_1 = {1 \over 2}\int \sqrt{-g} \left( {\pi^2\over
9\cdot 2^6}\, t_8 t_8 R^4 + {1\over 24} {E}_8 \right) -  (2\pi)^2 C_3 \wedge
X_8(R) +\ldots
\ee where $E_8$ is the $8$-dimensional Euler
density, normalized so that $\chi = \int d^8x \sqrt{g} E_8$. The
$8$-form, $X_8$, is a combination of the first and second Pontryagin
classes:
 \be \label{x8}
 X_8 = \frac{1}{192} (p_1^2 - 4p_2).
 \ee
This is normalized so that $\int_{\cM_8} X_8 = -{\chi(\cM_8)\over 24}$ when $\cM_8$ is complex. The $C_3\wedge X_8$ coupling contributes to the $G$ equation of
motion \bea d\star G + \half G \w G = -(2\pi)^2 X_8(R). \eea
The Pontryagin classes are given by
\be p_1 =
-\frac{1}{8\pi^2} \tr R^2,  \qquad p_2 = - \frac{1}{64\pi^4} \tr R^4
+ \frac{1}{128\pi^4} (\tr R^2)^2. \ee

In the context of M-theory compactified on an $8$-dimensional space,  these
known couplings are sufficient to evade the supergravity constraints
analogous to those presented in section~\ref{nogo}. With these
higher derivative couplings, compact flux compactifications are
possible~\cite{Becker:1996gj}. The same couplings permit four-dimensional F-theory
compactifications with flux, including the particular case of type IIB orientifolds~\cite{Sethi:1996es, Dasgupta:1999ss}.

We might first imagine that~\C{oldcouplings}\ might suffice to permit flux for M-theory on a $7$-manifold $\cM_7$. For example, one can orient $X_8$ along space-time and along a $4$-cycle of $\cM_7$. If the connection used to evaluate~\C{x8}\ is not just the spin connection $\omega$ but involves $G$-flux, one could imagine a membrane charge tadpole generated by $X_8$ in the presence of space-filling $G$-flux. The flux dependence of the connection would have to be something analogous to,
\be
\Omega_+ = \omega + {1\over 2} {\cal H}_3, \ee
used in the heterotic string to evaluate the Bianchi identity~\C{hetbianchi}. However, such a charge is not robust because space-time is topologically trivial and because the space-time $G$-flux is not quantized. Rather, it appears new couplings are needed beyond those of~\C{oldcouplings}, whose existence we will infer from duality.

In principle, we might worry that terms with more than $8$ derivatives might play a
role in permitting fluxes on a $7$-dimensional space; however, in all other
examples, that has not been the case. The obstruction is usually a
Gauss law constraint and the gravitational contribution to the
charge only comes at a fixed order in the momentum expansion, or at one-loop in a string loop expansion.
%We will see later whether this remains true in this case.

\subsection{Dualizing}

\label{mdualizing}

We need some new ingredient from the higher momentum interactions in the M-theory effective action. In four-dimensional type IIB flux compactifications, that ingredient was a D3-brane charge tadpole induced from four derivative gravitational couplings on $(p,q)$ $7$-branes proportional to,
\be\label{induced1}
\int C_4 \wedge p_1.
\ee
This coupling in type IIB is a consequence of the M-theory coupling,
\be\label{induced2}
\int C_3 \w X_8,
\ee
in a way that we will describe later. To find the new M-theory couplings, let us dualize~\C{induced2}. To dualize, we require some tools for computing curvatures on spaces with $U(1)$ isometries to which we now turn.

\subsubsection{Integrating out isometry directions}\label{integratingout}

The kind of coupling we need  will be at least quadratic in fluxes, yet it must generate a charge tadpole. Such couplings have not really made an appearance in string theory. To understand the structure of these couplings, we begin by reducing Pontryagin classes like $p_1$ of~\C{induced1}\ on spaces with  $U(1)$  isometries.

Consider a metric with an isometry direction parametrized by coordinate $y$,
\be\label{mmetric}
ds^2 = e^m e^m + e^y e^y, \qquad e^y = f(x) (dy + A).
\ee
We have chosen an orthonormal frame $e^m$ for the base of the circle fibration and coordinates $x$ for the base. The connection $1$-form $A$ describes the twisting of the circle over the base. Let $\omega^{mn}$ denote the spin connection for the base (in the absence of $e^y e^y$ terms) satisfying
\be
de^n + \omega^{nm} e^m =0.
\ee
Let us evaluate how the spin connection changes when the fibered circle is included. First define a $1$-form $g$ and a $2$-form $h$ via
\be\label{defineh}
 d (\log f) =g_n e^n,\qquad  f dA= h_{mn} e^m e^n.
\ee
The  components of the spin connection for the circle bundle take the form,
\be\label{spin1}
{\hat \omega}^{yn} = e^y g^n - h^{pn} e^p = -  {\hat \omega}^{ny},
\ee
and
\be\label{spin2}
{\hat \omega}^{nm} = {\omega}^{nm} - h^{nm} e^y.
\ee
The curvature two-forms are as usual:
\be
{\hat R}_{mn} = d {\hat \omega}_{mn} + {\hat \omega}_{mp} \wedge {\hat \omega}^p_{~~n}.
\ee
In computing quantities like $p_1$ on a Taub-NUT space, for example, we will meet expressions like
\be\label{explicitpont}
{4\pi^2} \, p_1 = R_{12} R_{12} + R_{13} R_{13} + R_{23} R_{23} + R_{1y} R_{1y} + R_{2y} R_{2y} + R_{3y} R_{3y}.
\ee
Any non-vanishing term on the right hand side of~\C{explicitpont}\ contains a single $dy$ factor, which implies a single $e^y$ factor using the orthonormal basis. This means that the connection $A$ never appears in this expression; only the field strength $dA$ appears via $h_{mn}$ of~\C{defineh}.

The explicit expressions for the curvatures of a circle bundle are given by
\bea
{\hat R}_{mn} &=&R_{mn} + d(h_{mn} e^y) - h_{mp} e^y \wedge \omega^p_{~~n} - \omega_{mp} \wedge h^p_{~~n} e^y + \cr
&& + g_m h_{np} e^p \w e^y - g_n h_{mp} e^p \w e^y - h_{qm} h_{pn} e^q \w e^p, \cr
&=& R_{mn} + h_{mn} h_{qp} e^q \w e^p -  h_{qm} h_{pn} e^q \w e^p \\
&& + \left( dh_{mn} + h_{mn} g^p e^p + h_{mp} \omega_{pn} - h_{np} \omega_{pm}  + g_m h_{np} e^p - g_{n} h_{mp} \right) \w e^y, \non
\eea
and
\bea
{\hat R}_{yn} &=& \left( {\hat \omega}_{py} e^p g_n - e_y dg_n   \right) + \left( e_y g_p - h_{kp} e^k \right) \wedge \left( \omega^p_{~~n} - h^p_{~~n} e_y\right) - d(h_{pn} e^p) \cr
&= & g_n h_{qp} e^q \w e^p  + d h_{np} e^p +  \omega_{np} h_{pq} e^q - h_{np} \omega_{pq} e^q  \cr
&& +  \left( dg_n - g_p \om_{pn} +  h_{np} h_{pq} e^q + g_n g_p e^p \right) \w e^y,
\eea
where we have separated out the terms proportional to $e^y$ for later convenience.

%\subsubsection{A torus fibration}
%The second case of interest to us is the case of a metric admitting a torus fibration with real coordinates $(w_1,w_2)$
%\be\label{mmetric}
%ds^2 = e^m e^m + e^w \bew, \qquad e^w = {1\over \sqrt{\tau_2}} (dw_1 + \tau dw_2), \qquad \tau=\tau(x).
%\ee

\subsubsection{Reducing M-theory to type IIA}

As an example, we can apply this formalism to M-theory reduced to type IIA string theory on~\C{mmetric}. We identify $A$ with the RR potential $C_1$ and $f=e^{4\phi/3}$ where $\phi$ is the dilaton. Our starting point is the coupling,
\be \label{gravcs}
\int C_3 \wedge X_8,
\ee
which reduces to,
\be \label{d6charge}
\int C_3 \wedge X_7,
\ee
on integration over $y$, where $X_7$ is constructed from metrics, curvatures, the field strength $F_2=dC_1$ and the dilaton $\phi$ in the way that we have described.

\subsubsection{An application to brane couplings}\label{branecouplings}

Before we get into the fairly complex $8$ derivative M-theory couplings related to $X_8$, let us turn to the simpler case of a $D7$-brane wrapping a surface $\M$ with a non-trivial circle bundle, for example, a Taub-NUT space. There is an induced $D3$-brane charge from $p_1(\M)$ of~\C{induced1}. We will T-dualize using the circle isometry which will give us a background with NS5-brane charge rather than gravitational charge. Tracking what happens to~\C{induced1}\ should provide us with a strong hint about the kind of coupling we desire in M-theory.

It is very useful to keep the example of a Taub-NUT space in mind. The Taub-NUT metric can be expressed in terms of coordinates $(r,\theta,\psi, y)$,
\be\label{TNmetric}
ds^2 =  V(dr^2 + r^2 d\Omega^2) + V^{-1}(dy + A )^2,
\ee
where
\be
V= 1 + {1\over r},  \qquad A=\cos\theta d\psi, \qquad d\Omega^2=d\theta^2 + \sin^2\theta d\psi^2.
\ee
The isometry direction  is the $y$-direction along which we dualize. After T-duality along $y$, the resulting metric, $B$-field and dilaton are given by
\be\label{dualTN}
{\widetilde ds^2} = V(dr^2 + r^2 d\Omega^2 + dy^2), \qquad B_{\psi y} = \cos\theta, \quad e^{2\phi} = V.
\ee
The space-time RR potential $(C_4)_{\mu_0 \ldots \mu_3} \rightarrow (C_5)_{\mu_0 \ldots \mu_3 y}$.  The background~\C{dualTN}\ is conformally ${\mathbb R}^3\times S^1$ but it supports $H$-flux on $S^2\times S^1$, which is the dual of the Kaluza-Klein charge of Taub-NUT. It describes a smeared NS5-brane. This is standard closed string T-duality.

We would like to rewrite $\tr (R\w R)$ evaluated with the Taub-NUT metric~\C{TNmetric}\ in terms of the dual variables~\C{dualTN}.  In terms of the formalism of the section~\ref{integratingout}, we identify
\be
f^2 = V^{-1}, \qquad g_r e^r = - {1\over 2} d(\log V), \qquad h_{mn} = f H_{mny},
\ee
where $H=dB$. Only $H$ will appear in the final expression as we explained in section~\ref{integratingout}. The indices $(m,n)$ refer to the orthonormal frame while $y$ is still a coordinate index. So we  need to evaluate:
\be
\tr (R\w R) = {\hat R}_{mn} \w {\hat R}_{nm}+{\hat R}_{yn} \w {\hat R}_{ny}.
\ee
We could simplify this expression by making use of the self-duality of the curvature $2$-forms for a $4$-dimensional hyperK\"ahler space like Taub-NUT if we desired, but we would like to see how this expression looks for a general surface with an isometry.

Unfortunately, even this  case is a little involved. Let us introduce some notation for $1$-forms which appear in the curvature expressions multiplying $dy$:
\bea\label{definealpha}
 \alpha_{mny} &=&  \left( dH_{mny} + 2 H_{mny} g^p e^p + H_{mpy} \omega_{pn} - H_{npy} \omega_{pm}  +  g_m H_{npy} e^p - g_{n} H_{mpy} \right), \cr
 \beta_n &=&  \left( dg_n - g_p \om_{pn} +  f^2 H_{npy} H_{pqy} e^q + g_n g_p e^p \right).
\eea
There are terms of at most quadratic order in the $H$-flux appearing in~\C{definealpha}. Evaluating $X_3^y=\int_y \tr (R\w R)$ in terms of the original Taub-NUT metric gives a $3$-form,
\bea
\int_y \tr (R\w R) = -  \left( R_{mn} + f^2 H_{mny} H_{qpy} e^q \w e^p - f^2 H_{qmy} H_{pny} e^q \w e^p \right) f^2 \alpha_{mny} -\\
  \left( g_n H_{qpy} e^q \w e^p  + g_q H_{npy} e^q\w e^p + d H_{npy} e^p +  \omega_{np} H_{pqy} e^q - H_{npy} \omega_{pq} e^q \right) f^2 \beta_n. \non
\eea
In terms of the dual metric~\C{dualTN}, we identify $f^2 = {\widetilde g}^{yy}$ giving a nicer expression
\bea
X_3^y &=& -  \left( R_{mn} + H_{mn}^{\phantom{mn}y} H_{qpy} e^q \w e^p - H_{qm}^{\phantom{qm}y} H_{pny} e^q \w e^p \right) \alpha_{mn}^y \\
&&  -\left( g_n H_{qp}^{\phantom{qp}y} e^q \w e^p  + g_q H_{np}^{\phantom{np}y} e^q\w e^p + d H_{np}^{\phantom{np}y} e^p +  \omega_{np} H_{pq}^{\phantom{pq}y} e^q - H_{np}^{\phantom{np}y} \omega_{pq} e^q \right) \beta_n. \non
\eea
We are not quite finished with expressing $X_3^y$ in T-dual variables. The inversion of $g_{yy}$ under T-duality means that ${\widetilde g}_n = -g_n$. We can finally express $X_3^y$  in terms of the $1$-forms,
\bea\label{defineoneforms}
 \alpha_{mny} &=&  \left( dH_{mny} - 2 H_{mny} {\widetilde g}^p e^p + H_{mpy} \omega_{pn} - H_{npy} \omega_{pm}  -  {\widetilde g}_m H_{npy} e^p + {\widetilde g}_{n} H_{mpy} \right), \cr
 \beta_n &=&  \left(  {\widetilde g}_p \om_{pn} - d{\widetilde g}_n +  H_{np}^{\phantom{np}y} H_{pqy} e^q + {\widetilde g}_n {\widetilde g}_p e^p \right),
\eea
where
\bea \label{dualcoupling}
X_3^y &=& -  \left( R_{mn} + H_{mn}^{\phantom{mn}y} H_{qpy} e^q \w e^p - H_{qm}^{\phantom{qm}y} H_{pny} e^q \w e^p \right) \alpha_{mn}^{\phantom{mn}y} \\
&&  + \left(  {\widetilde g}_n H_{qp}^{\phantom{qp}y} e^q \w e^p  +  {\widetilde g}_q H_{np}^{\phantom{np}y} e^q\w e^p - d H_{np}^{\phantom{np}y} e^p -  \omega_{np} H_{pq}^{\phantom{pq}y} e^q - H_{np}^{\phantom{np}y} \omega_{pq} e^q \right) \beta_n. \non
\eea
The proposed T-dual coupling is a $7$-form interaction supported on a D6-brane,
\be\label{d6coupling}
\int (C_5)_y \w X_3^y.
\ee
There is nothing special about the $y$-direction in this coupling; the covariant form of the coupling just involves a sum over all normal directions, $n^i$, to the D6-brane,
\be\label{covariantform}
\int (C_5)_{n^i} \w X_3^{n^i}.
\ee
This coupling is unusual because it involves an RR potential with legs normal to the brane. However, the coupling necessarily induces D4-brane charge by construction, even though the D4-brane is not supported on the D6-brane!

Note that $X_3^{n^i}$ involves terms linear and cubic in the flux $H$, and we have derived the coupling~\C{dualcoupling}\ without using any properties of Taub-NUT. From the perspective of this T-dual D6-brane, the coupling~\C{covariantform}\ is a world-volume $7$-form constructed from fluxes and potentials with legs  normal to the brane world-volume. It is an outstanding issue to recast~\C{covariantform}\ in a form that makes the geometry of $X_3$ more manifest. We will not pursue that question further here, though it is tied up with quite fascinating issues of Dirac quantization in the presence of fluxes.

Recently couplings of this general form involving one RR field, and one or two NS $B$-fields have been found on D-branes by other groups from string scattering computations~\cite{Garousi:2010ki, Garousi:2010bm, Becker:2011bw, Becker:2011ar, Hatefi:2012rx, Hatefi:2012ve, Hatefi:2012wj}, and from studying T-duality~\cite{Garousi:2009dj, Becker:2010ij, Garousi:2010rn, Garousi:2011ut, Garousi:2011fc, Velni:2012sv}.\footnote{These two collections of  interesting papers appeared after we had derived these couplings, independently of us. There is some overlap for the terms in~\C{dualcoupling}\ linear in $H$. Our derivation was presented at a number of conferences culminating in Strings 2010, Texas A\&M~\cite{strings2010}.} What is critical for us is that these couplings generate physical charge in the presence of flux.

\subsection{Lifting to M-theory}

At this point, we have learned about new couplings on D-branes that generate lower brane charge. Via T-duality, we expect such couplings to be present on all D-branes, not just D6-branes. Particularly for the case of D6-branes, however, it is natural to ask about the M-theory origins of these couplings. In doing so, we should learn about the new ingredients needed to evade the supergravity constraints of section~\ref{nogo}. The M-theory lift of the D6-brane couplings will be special in the sense that they can induce physical charge on a compact space.

%Now our interest is in learning about new M-theory couplings whose existence follows from~\C{gravcs}\ by duality.

The strategy goes as follows:  a D6-brane is a smooth Taub-NUT geometry in M-theory. On reduction to string theory, we can view the resulting background as either flat space with a D6-brane or the closed string D6-brane background. In either approach, further wrapping the D6-brane on a $4$-manifold with non-zero $p_1$  induces D2-brane charge. This charge arises from the brane supported coupling,
\be\label{d6branecoupling}
\int_{\rm D6} C_3 \wedge p_1,
\ee
or from $\int C_3\wedge X_7$ of~\C{d6charge}\ in the closed string approach. The coupling~\C{d6branecoupling}\ descends directly from $\int C_3 \wedge X_8$ of~\C{induced2}\ evaluated on Taub-NUT.  If we choose to replace Taub-NUT by an ALG space with an elliptic rather than circle fibration, we can further dualize to a type IIB D7-brane supporting a coupling proportional to $\int C_4 \wedge p_1$ of~\C{induced1}. Again we have both an open and closed string perspective.

The next step is to wrap the D7-brane on a circle-fibered $4$-manifold and dualize back to a type IIA D6-brane. From the open string brane perspective, this is the procedure described in section~\ref{branecouplings}. However, the closed string perspective provides a new coupling,
\be
\int C_5 \wedge X_5,
\ee
which induces D4-brane charge. Lifting this coupling to M-theory gives the new $8$ derivative couplings that produce M5-brane charge. In this duality chain, we are really considering M-theory on an $8$-dimensional space with a $T^3$-fibration and using T-duality to generate new couplings. There will be more couplings that can be discovered this way, but applying this chain to $\int C_3\wedge X_8$ will suffice to demonstrate new sources of M5-brane charge in four-dimensional M-theory flux compactifications.

Since we are concerned with higher derivative couplings, we need to worry about quantum corrections to the T-duality rules themselves. The existence of such corrections is very likely. However, we do expect the cohomology class of the induced charge to be captured using the standard uncorrected transformations which certainly transform brane charge correctly. The precise coupling might be shifted by exact terms, but our interest is really in the induced physical charge which is captured by the uncorrected rules.

To proceed, we take our ten-dimensional type IIA string metric to have the form,
\be\label{stringmetric}
ds^2 = e^m e^m + e^{y_1} e^{y_1} + e^{y_2} e^{y_2}, \qquad e^{y_i} = f_i(x) (dy_i + A_i(x)),
\ee
where the $y_i$ are space-like and the $e^m$ are independent of the $y_i$. We assume that the $1$-form $A_1$ has no component along $y_2$, while $A_2$ has components along $y_1$ and the $x$-directions. This is just a choice of parametrization for the metric~\C{stringmetric}. We will dualize along $(y_1, y_2)$ assuming $F_2=dC_1$ and the dilaton $\phi$ are independent of these coordinates so that T-duality can be applied. For convenience, we have summarized the T-duality transformations in Appendix~\ref{tduality}.
% If we were just dualizing to type IIB, it would be more natural to start with a torus fibered metric in M-theory which would make the dependence on the type IIB coupling $\tau$ manifest. Since we want to dualize back to type IIA, we may as well use~\C{stringmetric}.

We can simplify life by noting the initial coupling $\int C_3\wedge X_7$ of~\C{d6charge}\ does not require any $B$-field to generate charge.  Then, for simplicity we can start with a pure metric background~\C{stringmetric}\ with only a RR potential $C_1$ and dilaton. We could certainly consider a fully general background compatible with the assumed $T^3$-isometry, but that would complicate the resulting formulae.

Let us start by dualizing the $y_2$ direction.  The result is a type IIB background with metric, $B$-field and dilaton
\be
ds^2 = e^m e^m + e^{y_1} e^{y_1} +  \left( { dy_2 \over f_2} \right)^2, \qquad B=A_2, \qquad  e^{2 \phi_B} =  {e^{2 \phi}\over (f_2)^2}.
\ee
We can treat the dualization of the RR potentials separately since they do not affect the NS $B$-field, metric and dilaton.
The coupling~\C{d6charge}\ picks up several terms. The terms that interest us are going to be the ones that induce D3-brane charge. To find this T-dual coupling, let us consider $C_3$ oriented orthogonal to the $y_2$ direction then
\be\label{reduce}
\int C_3 \wedge X_7 \quad \rightarrow \quad\int C_3 \wedge X_6
\ee
after integration over $y_2$. This $9$-dimensional coupling must follow from reducing some type IIB coupling in ten dimensions which takes the form,
\be \label{iibcoupling}
\int C_4 \wedge X_6,
\ee
where $X_6$ depends on metrics, curvatures, $\phi_B$, and $C_0$ in the combination of the complexified string coupling $\tau_B$. This term is in the supersymmetric completion of the $R^4$ couplings in type IIB.
% It will lead to trivial shifts of the $D3$-brane charge in supergravity backgrounds.
In the background of a D7-brane, it will give rise to the $\int C_4 \wedge p_1$ coupling supported on the brane.
% in a way to be described in section~\ref{explicitd7coupling}.

The next step is to dualize along $y_1$. %This time we must allow a connection $A_1$.
This second T-duality takes us back to type IIA with background
\bea
&& ds^2 = e^m e^m + \left( { dy_2 \over f_2} \right)^2 + \left( {1 \over f_1} \right)^2 \left( dy_1 - (A_2)_{y_1} dy_2 \right)^2, \cr  && B_{my_1} =(A_1)_m, \qquad B_{my_2} = (A_2)_m - (A_2)_{y_1}(A_1)_m, \qquad  e^{2 \phi_A} =  {e^{2 \phi}\over (f_1 f_2)^2} .
\eea
We now apply the same logic to the coupling~\C{iibcoupling}; when reduced to $9$ dimensions, this coupling should arise, in part, from a type IIA coupling $\int C_5 \wedge X_5$. This $X_5$ is what we really seek. It is the closed string analogue of the $X_3$ coupling found in~\C{dualcoupling}. We also need to track the fate of the RR potential $C_1$ through this chain of two T-dualities. This is straightforward with the resulting primed RR potentials given by,
\bea
 & (C_1')_{y_1} = (C_1)_{y_2}, \qquad (C_1')_{y_2} =-(C_1)_{y_1}, & \cr
& (C_3')_{my_2y_1} =  (C_1)_m - (C_1)_{y_2} (A_2)_m  - (C_1)_{y_1} (A_1)_m  + (C_1)_{y_2} (A_1)_m  (A_2)_{y_1}. &
\eea

So far, our discussion is in terms of string frame variables. The final step is to recast the discussion in terms of M-theory variables. Start with an M-theory background with metric
\bea\label{startmetric}
 & ds^2 = e^m e^m  + e^{y_1} e^{y_1} + e^{y_2} e^{y_2}+ e^y e^y,  \qquad e^y = f(x) (dy + A(x)), &\cr
 &e^{y_1} = f_1(x) (dy_1 + A_1(x)), \qquad e^{y_2} = f_2(x) (dy_2+A_2(x)), &
\eea
and no flux. The potential $A$ has components in all directions, the potential $A_2$ has components in all directions except $y$, while the potential $A_1$ has components in directions except $y$ and $y_2$. Again, this is just a choice of parametrization. Reducing to type IIA gives a string-frame metric, dilaton and RR $1$-form potential:
\be\label{stringframeIIA}
ds^2 = f \left( e^m e^m  + e^{y_1} e^{y_1} + e^{y_2} e^{y_2} \right), \quad e^{\phi}  = f^{3/2} , \quad C_1= A.
\ee
We can run this IIA metric and $C_1$ through the duality chain above and lift back to M-theory to get a hatted $11$-dimensional metric and $3$-form potential ${\widehat C}_3$,
\be
{\widehat ds^2}= e^{\widehat m} e^{\widehat m}  + e^{\widehat y_1} e^{\widehat y_1} + e^{\widehat y_2} e^{\widehat y_2}+ e^{\widehat y} e^{\widehat y}, \quad {\widehat C}_3,
\ee
given by,
\bea
& {\hat f}^3 = {f\over (f_1 f_2)^2}, \qquad e^{\widehat m} = (f f_1 f_2)^{1/3} e^m, & \cr & e^{\widehat y} = {\hat f}(dy + \widehat{A})= {\hat f}(dy +(A)_{y2}dy_1 - (A)_{y_1} dy_2), &\cr
& e^{\widehat y_2} = { (f_1)^{1/3} \over (f f_2)^{2/3}} dy_2, \qquad e^{\widehat y_1} = { (f_2)^{1/3} \over (f f_1)^{2/3}} (dy_1 - (A_2)_{y_1} dy_2), & \cr & ({\widehat C}_3)_{my_1y} = (A_1)_{m}, \qquad ({\widehat C}_3)_{my_2y} = (A_2)_{m}- (A_2)_{y_1}(A_1)_m, & \cr &
({\widehat C}_3)_{my_2y_1} =  (A)_m - (A)_{y_2} (A_2)_m  - (A)_{y_1} (A_1)_m  + (A)_{y_2} (A_1)_m  (A_2)_{y_1}. &
\eea
This expresses all the data of the resulting M-theory metric and flux in terms of the original pure M-theory metric~\C{startmetric}.

This map is invertible, and it is the inverse expressing the original M-theory metric data in terms of hatted variables which is more useful. The inverse map for the diagonal metric components is given by,
\be\label{diagmap}
f^3  = { {\hat f} \over ({\hat f_1} {\hat f_2})^2}, \qquad f_1^3 = {{\hat f_2}\over ({\hat f} {\hat f_1})^2}, \qquad f_2^3 = {{\hat f_1}\over  ({\hat f} {\hat f_2})^2}, \qquad e^m = ({\hat f} {\hat f_1} {\hat f_2})^{1/3} e^{\widehat m},
\ee
where we identify $f^2=g_{yy}, {\hat f}^2 = {\hat g}_{yy}$ etc. The off-diagonal metric components follow from the relations,
\bea\label{offdiag}
&&(A_1)_m =  ({\widehat C}_3)_{my_1y}, \qquad (A_2)_{y_1} = - ({\widehat A_2})_{y_1}, \qquad (A_2)_{m} = ({\widehat C}_3)_{my_2y} -({\widehat A_2})_{y_1} ({\widehat C}_3)_{my_1y}, \cr && (A)_{y_1} = - ({\widehat A})_{y_2} ,\qquad (A)_{y_2} = ({\widehat A})_{y_1}, \cr && (A)_m = ({\widehat C}_3)_{my_2y_1} + ({\widehat A})_{y_1}  ({\widehat C}_3)_{my_2y}  - ({\widehat A})_{y_2} ({\widehat C}_3)_{my_1y}.
\eea
There is a nice cancelation in the final line of~\C{offdiag}\ so no terms quadratic in ${\widehat A}$ appear. 

Now we can finally describe the coupling $X_5$. Take $X_8$ evaluated on a $T^3$-fibered metric~\C{startmetric}\ and integrate over $(y,y_1,y_2)$ using the formulae for curvatures given in section~\ref{integratingout}:
\be
X_5 = \int dy \, dy_1 dy_2 \, X_8.
\ee
The resulting expression for $X_5$ is constructed from the metric and derivatives. Using the maps~\C{diagmap}\ and~\C{offdiag}, those metric components can be expressed in terms of the hatted metric and ${\widehat C_3}$-flux.

The result is a particular case of a covariant $8$ derivative M-theory coupling built from flux and the metric,
\be -{1\over 2} (2\pi)^2 \int C_6 \wedge X_5,
\ee
in the normalization of the couplings appearing in~\C{oldcouplings}.
It is a distinguished coupling because it can generate M5-brane charge on compact spaces. We can make the form of this coupling considerably nicer by noting that Pontryagin classes, and hence $X_8$, are conformally invariant.\footnote{There is a potential subtlety in this statement for non-compact spaces. For non-compact spaces, integrated characteristic classes are weakly metric-dependent, and one should be careful about making conformal transformations of the metric. Our eventual application of these couplings is to compact spaces so we will ignore this possible complication.} Up to the conformal factor $({\hat f} {\hat f_1} {\hat f_2})^{2/3}$, the metric used to evaluate $X_8$ takes the much cleaner form:
\be\label{simplifiedstory}
ds^2 = e^{\widehat m} e^{\widehat m}  + {1\over ({\hat f} {\hat f_1})^2} (dy_1+A_1)^2 + {1\over ({\hat f} {\hat f_2})^2} (dy_2+A_2)^2+ {1\over ({\hat f_1} {\hat f_2})^2} (dy+A)^2.
\ee
We can much more clearly see that the $G$-flux is modifying only the circle bundle connections $(A, A_1,A_2)$, which are expressed in terms of hatted variables in~\C{offdiag}.
Clearly, this coupling cries out for a more natural geometric interpretation, along with an analogous interpretation for the $1$-forms $\alpha$ and $\beta$ of~\C{defineoneforms}. Such an interpretation is likely to involve a better understanding of anomaly cancelation in the presence of fluxes.

\section{Examples}
\label{compact}

If there were no examples of backgrounds using the mechanism of section~\ref{dualizing}, it would be of limited interest. However, we can construct type IIA and M-theory backgrounds preserving both ${\cal N}=1$ and ${\cal N}=2$ supersymmetry by duality. Our starting vacua are, perhaps, the nicest examples of flux vacua; they are type IIB orientifolds of the form $K3\times {T^2\over  (-1)^{F_L} \Omega \, {\Z_2}}$ constructed in~\cite{Dasgupta:1999ss}. These backgrounds are special loci of more general F-theory compactifications on $K3\times {\widehat K3}$ with flux. On these special loci, each of the four $O7^-$-planes generated by the orientifold action has its charge canceled point-wise  by four D7-branes. This allows the type IIB string coupling to remain constant over the compactification manifold.

This geometric background preserves ${\cal N}=2$ supersymmetry. However, the choice of flux can preserve either ${\cal N}=2$ or ${\cal N}=1$ supersymmetry. The flux is actually most beautifully described in terms of M-theory data on $K3\times {\widehat K3}$. In this case, $[{G\over 2\pi}]$ is a primitive element of $H^{(2,2)}(K3\times {\widehat K3}, \Z)$. Such classes are either purely the wedge product of $(1,1)$ forms on both $K3$ surfaces, or include $\omega^{(2,0)} \wedge{\widehat \omega}^{(0,2)}$. The latter case preserves only  ${\cal N}=1$ supersymmetry. For a purely flux background with no branes, we demand that
\be
{1\over 2} \int {G\over 2\pi}\w {G\over 2\pi} = {{\chi(K3\times {\widehat K3})}\over 24}.
\ee
There are many choices of flux solving this tadpole constraint for both the ${\cal N}=1$ and ${\cal N}=2$ cases.

When lifted to type IIB, this $G$-flux becomes a combination of $F_3$ and $H_3$ fluxes. There is also an accompanying $F_5$ flux determined by the warp factor. Let $\alpha \in H^{1,1}(K3,\Z)$ and $\beta \in H^{2,0} (K3,\Z)$ be primitive classes. For simplicity, take a square canonical complex structure for $T^2$ with complex coordinate $z = z_1+iz_2$. The type IIB metric is given by,
\be\label{iibmetric}
ds^2 = e^{-3w /4} \eta_{\mu\nu} dx^\mu dx^\nu + e^{3w/4}( ds^2_{K3} + dz d{\bar z}),
\ee
with $w$ the warp factor, which depends on the internal coordinates. The fluxes are given by,
\bea \label{startingfluxes}
& H_3 = (\alpha +\betabar)\w dz + \cc, \qquad F_3 =dC_2= i(\betabar - \alpha) \w dz + \cc, &\cr
& F_5 = \ve_4 \w d e^{-3w/2} + H_3\w C_2,&
\eea
with $\ve_4$ the volume form for Minkowski space-time. The equation determining the warp factor is of Laplace type and follows from self-duality of $F_5$ together with the Bianchi identity,
\be
\Box e^{3w/2} = \star_6 \left( F_3 \w H_3 \right)+ S(X_8)\label{eqn:laplacian},
\ee
where $\star_6$ is with respect to the unwarped internal metric $ds^2_{K3} + dz d{\bar z}$ of~\C{iibmetric}. The term $S(X_8)$ labels the higher derivative contribution that makes a solution of~\C{eqn:laplacian}\ possible on a compact space. In M-theory, this source is $C_3\wedge X_8$ evaluated on $K3\times {\widehat K3}$. In type IIB, the source of D3-brane charge is the $C_4\wedge p_1$ coupling supported on the $O7$-planes and D7-branes.

One can imagine dualizing this background in many ways. Dualizing along the $T^2$, which is quotiented by the action $(-1)^{F_L} \Omega\, {\Z_2}$, gives rise to torsional type I and heterotic solutions described in~\cite{Dasgupta:1999ss}; these backgrounds, which involve ``geometric flux,'' have been the subject of much study.\footnote{The terminology ``geometric flux" is a colloquial (and somewhat inappropriate) way to describe a circle bundle that results from $H$-flux becoming metric after T-duality. Essentially, reversing the duality between a Taub-NUT space and a smeared NS5-brane described in section~\ref{branecouplings}. This duality first appeared in flux compactifications in~\cite{Dasgupta:1999ss}, and was also discussed in massive supergravity in~\cite{Kaloper:1999yr}. The problem with this terminology is that it suggests a symmetric treatment of conventional and geometric flux. Particularly in constructing flux superpotentials for four-dimensional effective field theories. However, these two notions are physically distinct at large volume. Changing geometric flux changes the topology of the compactification manifold. The collection of light scalar fields to be included in any four-dimensional effective theory depends on the topology of the vacuum manifold. Unlike conventional flux, whose effects can be described by a superpotential for large volume compactifications, each choice of ``geometric flux" typically corresponds to a different low-energy theory. In a fully quantum treatment of $H$-flux via string theory rather than supergravity, it should be the case that both geometric and conventional flux appear on similar footing since they are related by a  perturbative string duality. This equivalence is captured by $H$-twisted versions of cohomology and K-theory; see, for example~\cite{Bergman:2007qq}.} One can also choose a special $K3$ metric and dualize along the $K3$ surface. Depending on the choice of dualization, non-geometric heterotic, type IIB and M-theory backgrounds can be found~\cite{Becker:2009df, McOrist:2010jw}; see, for example,~\cite{Shelton:2005cf}\ for a discussion of non-geometric backgrounds constructed by T-dualizing $H$-flux.

We want to follow the chain of reasoning presented in section~\ref{dualizing}. To follow that chain requires a single T-duality along the $K3$ direction taking us from type IIB with D7-branes to type IIA with D6-branes. Smooth $K3$ surfaces have no exact $U(1)$ isometries, but we can either take an orbifold or semi-flat metric for the $K3$ surface, or replace $K3$ by Taub-NUT if one prefers studying a local smooth model. Since we discussed the local case in section~\ref{dualizing}, let us consider an compact elliptic $K3$ surface either of the form $T^4/\Z_2$, or with a semi-flat approximation to the exact smooth metric:
\be\label{semiflat}
ds^2_{K3} = g(u) du d{\bar u} + {1\over \tau_2(u)}| dv_1 + \tau(u) dv_2|^2.
\ee
The coordinate $u$ parametrizes the $\PP^1$ base of the elliptic fibration for $K3$ while $(v_1, v_2)$ parametrize the fiber torus. There is a difference between using a semi-flat metric and an orbifold metric for $T^4/\Z_2$. For a nice square case, the orbifold metric would simply be the flat metric for $T^2\times T^2$,
\be\label{orbifold}
ds^2_{\rm orbifold} = du d{\bar u} + dv d{\bar v}, \qquad v = v_1+iv_2,
\ee
orbifolded by the action  $(u,v) \rightarrow (-u,-v)$. Combining the orbifold action with the orientifold action  ${T^2\over  (-1)^{F_L} \Omega \, {\Z_2}}$ generates both $O7$-planes and $O3$-planes. The presence of $64$ $O3$-planes changes the tadpole condition in a manner that depends on the number of $O3^{-}$-planes versus $O3$-planes of other flavors. For example, if all the $O3$-planes were $O3^-$-planes the tadpole condition would require an additional $16$ units of D3-brane charge from either branes or fluxes. We really want to use the orbifold metric as a convenient approximation to a smooth $K3$ metric so the presence of these $O3$-planes is really just a distraction.

A single T-duality along the elliptic fiber of the $K3$ surface, say along $v_1$, will produce both $O6$-planes and $O4$-planes. This is fine for describing a type IIA flux solution. However, the lift to M-theory will be an M-theory orientifold, which involves the inversion
\be\label{morientifold}
C_3 \, \rightarrow - C_3,
\ee
as well as a geometric quotient. The extra orientifold action~\C{morientifold}\ comes from the strong coupling description of $O4$-planes. We will revisit this point in section~\ref{mliftsection}. %If we want a purely geometric M-theory lift, we are better off studying the semi-flat case.

Once again, the metric that results from T-duality only cares about the initial NS sector data consisting of the starting metric~\C{semiflat}\ and the $H_3$-field of~\C{startingfluxes}. We do need a local potential for $H_3$ and the natural choice is to consider,
\be\label{defineB}
B_2 = z (\alpha +\betabar) + \cc,
\ee
trivializing along the ${T^2\over  (-1)^{F_L} \Omega \, {\Z_2}}$ factor. Now let us examine particular cases with the aim of unraveling the essential structure.

\subsection{Orbifold case}\label{orbifoldsection}

The simplest case to treat is the orbifold metric~\C{orbifold}. In order to T-dualize along the $v_1$ direction, we need to decompose $B_2$ of~\C{defineB}\ as follows,
\be\label{decomposeB}
B_2 = {\widetilde B_2} + {\widetilde A} dv_1,
\ee
where ${\widetilde A}$ is a $1$-form connection and  ${\widetilde B_2}$ has no $dv_1$ component. The resulting type IIA  metric  given by,
\be\label{tdualexample}
ds^2_{\rm IIA} = e^{-3w /4} \eta_{\mu\nu} dx^\mu dx^\nu +  e^{3w/4}( dz d{\bar z} + du d{\bar u} + (dv_2)^2)  + e^{- 3w/4}(dv_1 + {\widetilde A})^2,
\ee
now involves a non-trivial circle bundle, or equivalently, ``geometric flux." The orientifold action is now generated by $(z,v_1) \rightarrow (-z,-v_1)$ coupled with $\Omega (-1)^{F_L}$.
% and by the geometric quotient $(u, v_1, v_2) \rightarrow (-u,-v_1, -v_2)$ coupled with $(-1)^{F_L}$.
The IIA $B$-field and dilaton take the form,
\be
{\widetilde B_2}, \qquad  \phi_{\rm IIA} = \phi - {3w\over 8}, \ee
where the type IIB dilaton, $\phi$, is a constant for the initial orientifold compactification.
 To specify the RR potentials, we need to decompose $C_2$ in a similar way:
\be\label{decomposeC}
C_2 = iz (\betabar - \alpha) + {\rm c.c.} = {\widetilde C}_2 + {\widetilde C}_1 dv_1.
\ee
The IIA $1$-form and $3$-form RR potentials are given by
\be
({\widetilde C}_1)_i, \qquad ({\widetilde C}_2)_{ij v_1}, \qquad 3 ({\widetilde C}_2)_{[ij} ({\widetilde A})_{k]}.
\ee
What is important is that this combination of fluxes together with the geometry~\C{tdualexample}, and the assembled $O6$/D6 system induces a D4-brane charge tadpole which permits a Minkowski space-time.

\subsection{The semi-flat case}\label{semi}

The semi-flat metric~\C{semiflat}\ is a very good approximation to the exact $K3$ metric. This metric deviates from the smooth $K3$ metric only in a very small neighborhood of each degeneration of the elliptic fiber; for a very accurate approximation to the $K3$ metric built by repairing the semi-flat metric, see~\cite{gross-2000}. Dualizing the semi-flat metric along $v_1$ is only slightly more involved than the orbifold metric. The resulting type IIA metric takes the form
\bea\label{semiflatmetric}
ds^2_{\rm IIA} &=& e^{-3w /4} \eta_{\mu\nu} dx^\mu dx^\nu +  e^{3w/4} \left( dz d{\bar z} + g(u) du d{\bar u} + \tau_2 (dv_2)^2 \right)
\cr && + e^{- 3w/4} \tau_2 (dv_1 + {\widetilde A})^2.
\eea
The $\tau$ monodromies of the elliptic fiber have become $\rho$ monodromies. In the neighborhood of a singular fiber, the metric must be repaired if we want a smooth background.
Using the same decomposition~\C{decomposeB}, the type IIA $B$-fields and dilaton are given by,
\be
 (\widetilde{B_2})_{ij} +  \tau_1 {\widetilde A}_j \delta_{i,v_2}  \quad i,j\neq v_1, \qquad (\widetilde{B_2})_{v_2v_1} = \tau_1, \qquad \phi_{\rm IIA} = \phi - {3w\over 8}+ {1\over 2} \log \tau_2.
\ee
The type IIA RR $1$-form and $3$-form potentials are given by,
\be ({\widetilde C}_1)_i, \qquad ({\widetilde C}_2)_{ij v_1}  - \tau_1 ({\widetilde C}_1)_i \delta_{j,v_2}, \qquad 3 ({\widetilde C}_2)_{[ij} ({\widetilde A})_{k]} + 2 \tau_1 ({\widetilde C}_1)_{[i} ({\widetilde A})_{j]} \delta_{k,v_2 }.
\ee

There is a further interesting duality worth mentioning at this point. If we set $\beta$ of~\C{startingfluxes}\ to zero then ${\cal N}=2$ supersymmetry is preserved. In this case, there is a further duality relating  F-theory on $K3\times {\widehat K3}$ and other IIB orientifolds to compactifications of the type IIA string on conventional $CY_3$ spaces.  This duality has been explored in cases where the anomaly is canceled purely by branes~\cite{Sethi:1996es}, and by combinations of branes and fluxes~\cite{Schulz:2004tt, Schulz:2012wu, Melnikov:2012cv}. Now we are presenting a further duality to type IIA with fluxes. This suggests a IIA/IIA duality between a class of Calabi-Yau compactifications and flux vacua. Tracking the string coupling and volume factors through this duality chain might provide a new computational approach for determining the quantum corrected vector and hypermultiplet moduli spaces, along the lines of~\cite{Halmagyi:2007wi}. We will not pursue this direction here, but rather turn to the M-theory lift of these vacua.

\subsection{M-theory lift} \label{mliftsection}

Let us lift the orbifold solution of section~\ref{orbifoldsection}\ to M-theory. We again use $y$ to label the circle taking us from M-theory to type IIA. To simplify the solution, we will shift $w$ to absorb the constant IIB dilaton $\phi$. This is a choice that involves rescaling the space-time coordinates $x^\mu$. The M-theory metric then takes the form
\bea\label{mlift}
ds^2_{\rm M} &=& e^{ -w /2} \eta_{\mu\nu} dx^\mu dx^\nu +  e^{w} \left( dz d{\bar z} + du d{\bar u} +  (dv_2)^2 \right)
\cr && + e^{- w/2}  (dv_1 + {\widetilde A})^2 + e^{-w/2} (dy + {\widetilde C}_1)^2.
\eea
It is pleasing that $v_1$ and $y$ appear on symmetric footing with one circle bundle determined by the NS $B_2$-field via~\C{decomposeB}, and one determined by the RR $C_2$-field via~\C{decomposeC}. The basic structure of~\C{mlift}\ is a torus bundle over $T^5$.  The M-theory $3$-form ${\widehat C}_3$ takes the form,
\be\label{mfluxes}
({\widehat C}_3)_{ijy} = ({\widetilde B_2})_{ij}, \qquad  ({\widehat C}_3)_{ijv_1} = ({\widetilde C}_2)_{ij v_1}, \qquad   ({\widehat C}_3)_{ijk} = 3 ({\widetilde C}_2)_{[ij} ({\widetilde A})_{k]}.
\ee
The tilde IIA fluxes are all linear in $z$ with $i,j,k$ indices in the $(u,v)$ directions. There is an additional orbifold action generated by two elements. The first sends
\be\label{firstmquotient}
(z, v_1, y) \rightarrow (-z, -v_1, -y),
\ee
again treating $v_1$ and $y$ symmetrically. This action is the lift of the type IIA orientifold action. Note that the M-theory fluxes~\C{mfluxes}\ are invariant under this action. The second generator is the image of the $\Z_2$ generator used to construct the orbifold $T^4/\Z_2$ in type IIB. By tracking the action of this generator on the RR fields, we see that it corresponds to an orientifold action in M-theory sending,
\be\label{morientifold}
(u, v_1, v_2, y) \rightarrow (-u,-v_1, -v_2, -y), \qquad {\widehat C}_3 \rightarrow - {\widehat C}_3.
\ee
It is the inversion of ${\widehat C}_3$ that makes this an orientifold action. This is quite natural since we are lifting a background with $O4$-planes. Each $O4$-plane has a local description in M-theory as an orientation reversing orbifold ${\mathbb R}^5/\Z_2 \times S^1$ where the $\Z_2$ also inverts ${\widehat C}_3$. Note that the fluxes~\C{mfluxes}\ are all odd under the geometric action~\C{morientifold}. Another way to see this orientifold action is by noting that IIB on $T^4/\Z_2$ maps to IIA on $T^4/\Z_2 (-1)^{F_L}$ under a single T-duality.

If we had started simply with type IIB on $K3$ with no additional torus or fluxes then there is a duality relating
\be
\mbox{M-theory} \quad T^5/\Z_2 \quad \Leftrightarrow \quad \mbox{IIB} \quad K3,
\ee
which essentially follows from this same chain of manipulations~\cite{Dasgupta:1995zm, Witten:1995em}. What we have found is an extension of this duality relating type IIB orientifolds of $T^2\times K3$ with flux to M-theory flux vacua.

To a large extent, the additional structure from lifting $O4$-planes is a distraction. If we could have dualized a smooth $K3$ metric, there would be no $O4$-planes and corresponding $\Z_2$ action~\C{morientifold}, but there would be a purely geometric background. For example, we could approximate a $K3$ metric locally by a smooth Taub-NUT space and follow it precisely through this chain. This is essentially what we did in section~\ref{dualizing}. That really makes the existence of smooth compact M-theory flux backgrounds preserving ${\cal N}=1$ and ${\cal N}=2$ supersymmetry dual to type IIB orientifolds of $T^2\times K3$ highly likely. Proving the existence of such backgrounds is, however, likely to be a very non-trivial problem.

There are a number of variants of this construction. For example, we could replace the $O4^-$-planes with $O4^{-'}$-planes which correspond to a single D4-brane stuck to each orientifold plane. Each $O4^{-'}$-plane carries no net D4-brane charge. The M-theory lift of this plane is ${\R^5 \times S^1}/\Z_2$ where the $\Z_2$ acts freely with a $1/2$ shift along the M-theory circle~\cite{Hori:1998iv}. If we normalize $y$ to have period $1$ then the M-theory orientifold action is generated by~\C{firstmquotient}\ together with
\be\label{morientifoldtwo}
(z, u, v_2, y) \rightarrow (-z,-u, -v_2, y+{1\over2}), \qquad {\widehat C}_3 \rightarrow - {\widehat C}_3.
\ee

%We can get a better feel for what such a solution might look like by lifting the semi-flat type IIA metric of~\C{semiflatmetric}\ to M-theory. This should give a very good approximation to the smooth solution away from the discriminant locus of the $K3$ surface.

%\subsection{Local supersymmetry constraints}
%\label{gstructure}

\section{Massive type IIA Supergravity}\label{massive}

The final topic of discussion is type IIA with a Romans mass~\cite{Romans:1985tz}. The mechanism we  described for M-theory and type IIA flux compactifications is quite similar to the mechanisms allowing flux in type IIB  and heterotic string theory. There is one other proposed mechanism for flux compactifications in massive type IIA supergravity  by DeWolfe et. al.~\cite{DeWolfe:2005uu}. Starting with a Calabi-Yau geometry, those authors appear to find a very striking class of ${\cal N}=1$ supersymmetric $AdS_4$ compactifications with an internal volume that can be made large and a  string coupling that can be made small.

Indeed, the string coupling can be made parametrically small, while the separation of the $AdS_4$ scale from the compactification scale can be made parametrically large. The parameter corresponds to the amount of internal $G$-flux.
The only ingredient in these compactifications beyond massive IIA supergravity are $O6$-planes. The basic idea is to cancel the negative charge of each $O6$-plane not with D6-branes, but with $H_3$-flux. In the presence of a Romans mass, $H_3$-flux sources D6-brane charge.
If the construction is valid, this is a very striking family of solutions exhibiting properties not seen in any other known construction of flux vacua. It would sharply differentiate massive IIA from conventional string theory or M-theory.

However, there are reasons to be uneasy about the proposal of~\cite{DeWolfe:2005uu}. The authors start with a Calabi-Yau background and consider the effects of flux. This is a reasonable approach in type IIB string theory where the flux back-reaction alters the metric, but not the topological type of the compactification. It is not a reasonable approach in heterotic string theory where flux vacua are non-K\"ahler manifolds topologically distinct from Calabi-Yau spaces. In supergravity, one must choose a manifold of fixed topological type together with flux of fixed topological type, and study whether there is a solution to the equations of motion as parameters of the metric are varied. In string theory, it is possible to change topological type with finite energy but this cannot happen in supergravity.

In the large volume limit, the energy of localized fluxes like $G$ or $H_3$ becomes less important. In type IIB string theory, this means a large volume solution approaches Calabi-Yau. This is not true for the Romans theory because the Romans parameter $m$, or $F_0$ when viewed as an RR flux, does not dilute.  Indeed $AdS_4\times CY_3$ is not an approximate solution of massive IIA, and it is unclear why an expansion around a Calabi-Yau space is a sensible starting point.

The second cause of concern is the addition of an orientifold plane. Massive IIA has no quantum description, similar to a perturbative string expansion, which could be used to define an orientifold. This worry is not uniquely ours, but has been explored in~\cite{Banks:2006hg, moore-unpublished}. As we explained in section~\ref{o6enough}, a correct definition of  $O6$-planes is subtle and is correlated with  the parity of the Romans parameter.  There is an interesting recent attempt to study an $O6$-plane in massive IIA which is partly analytic and partly numerical~\cite{Saracco:2012wc}. The end result of the analysis appears to be an orientifolded geometry that is regular in massive IIA supergravity, without any exotic ingredients.

This background itself may well be a fine solution of massive IIA, but it is unlikely to describe an orientifold with negative tension. Orientifolds are typically singular solutions in supergravity requiring either a change in dimension (like the case of $O6$-planes in conventional IIA), or other strong stringy effects to desingularize the physics. In asymptotically flat space-time, this is a requirement of the positive mass theorem since $O6$-planes can be viewed as particles with negative mass in the spatially transverse $3$ dimensions. It is intuitively hard to see how a negative tension object could have a regular metric. The analysis of~\cite{Saracco:2012wc}\ suggests that an $O6$-plane becomes a regular background with bounded dilaton in massive IIA, which would not differentiate it from a conventional source of stress-energy. In particular, it would not help evade any no-go theorems. Regardless, it is fair to say that more insight is needed to understand whether $O6$-planes with negative tension exist in massive IIA.

As one can see, there are several poorly understood ingredients  involved in building the vacua of~\cite{DeWolfe:2005uu}. Let us take a step back from the complications; we will ignore concerns about defining $O6$-planes, or problems with expanding around a non-solution like a Calabi-Yau metric. All of these issues are forgivable if flux vacua with such striking features exist. Rather, let us see what can be said directly from examining the equations of motion. As a preliminary, we note that a no-go result along the lines of section~\ref{nogo}\  forbidding  Minkowski and de Sitter solutions in massive type IIA was nicely described in~\cite{Maldacena:2000mw}.

The fields of massive type IIA consist of a metric $g$, $4$-form flux $G$, $3$-form NS flux $H=dB$, $2$-form RR flux $F$, and dilaton $\phi$. The Romans mass parameter is $m$. We will follow the conventions of~\cite{Lust:2004ig, Acharya:2006ne}. These conventions are also used  in the attempt by~\cite{Acharya:2006ne}\ to build ten-dimensional solutions of the type proposed in~\cite{DeWolfe:2005uu}\ with smeared orientifold planes. As noted by various groups including~\cite{Acharya:2006ne, Saracco:2012wc, Banks:2006hg, moore-unpublished, Blaback:2010sj, Blaback:2012mu}\ along with us, smearing orientifolds is not sensible in string theory; we will only consider localized planes.

We again consider a warped $10$-dimensional metric of the form~\C{eqn:metric_1}\ with unwarped space-time and internal metrics: ${\hat g}^{(4)}_{\mu\nu}(x)$ and ${\hat g}^{(6)}_{mn}(y)$. We take $4$-form flux with the same ansatz~\C{eqn:four-form}. There are two global constraints that follow from the equations of motion for the scalars $\phi$ and $w$. Let us start with the dilaton equation of motion,
\bea\label{massivedilaton}
e^{-10w} {\hat \nabla}_m \left( e^{8w} {\hat g}^{mn} \partial_n \phi \right) &=& 5m^2 e^{5\phi/2} - {1\over 4} e^{\phi/2} f^2+{3\over 4} e^{3\phi/2} |F|^2  - 3\pi \sqrt{\alpha'} e^{3\phi/4} {1 \over \sqrt{g_3^t}} \delta^3(O6) \cr &&- {1\over 2} e^{-\phi} |H|^2 + {1\over 4} e^{\phi/2} |G^{\rm int}|^2,
\eea
where we have used the expression for the $O6$-plane source from~\cite{Acharya:2006ne}. The determinant of the metric transverse to the $O6$-plane is denoted $g_3^t$. The terms on the right hand side of~\C{massivedilaton}\ are arranged according to importance as the volume of the internal metric, ${\hat g}^{(6)}$, is scaled up. For example, the Romans term dominates at very large volume. The left hand side is written in a way that makes the global constraint manifest. The constraint is obtained by integrating~\C{massivedilaton}\ over the internal space with measure $e^{10w} \sqrt{{\hat g}^{(6)}}$.

The second constraint on $w$ follows from the space-time components of the Einstein equations,
\bea \label{massiveeinstein}
{\hat \cR}_{\mu\nu} &=& {\hat g}_{\mu\nu} e^{2w} \Bigg( e^{-2w} {\hat\nabla}^2 w+ 8 e^{-2w}  | {\hat\nabla} w|^2 - {1\over 2} e^{\phi/2} f_0^2 e^{-8w} - {3\over 16} e^{\phi/2} |G|^2  - {1\over 8} e^{-\phi} |H|^2  \cr &&  - {1\over 16} e^{3\phi/2} |F|^2 + {1\over 4} m^2 e^{5\phi/2} +{\pi\over 4} \sqrt{\alpha'} e^{3\phi/4} {1 \over \sqrt{g_3^t}} \delta^3(O6) \Bigg).
\eea
For $AdS_4$, the right hand side must again be independent of the $y$ coordinates. The global constraint comes from writing,
\be
{\hat\nabla}^2 w+ 8  | {\hat\nabla} w|^2 = {1\over 8} e^{-8w} {\hat \nabla}^2 e^{8w},
\ee
and again integrating over the internal space. For the moment, we will focus on the stronger point-wise constraint from~\C{massiveeinstein}.

Now we can turn to the solutions of~\cite{DeWolfe:2005uu}. The key observation is that $H$ magnetically sources $F$ in massive IIA,
\be\label{massiveH}
dF = 2mH  - 4\pi\sqrt{\alpha'} \delta^3(O6).
\ee
Using this supergravity source, additional D6-branes are not needed to cancel the charge of the $O6$-plane and condition~\C{massiveH}\ fixes the amount of $H$-flux. However, this cancelation is not point-wise. Because the right hand side involves canceling a point source against a smooth charge distribution in the three directions transverse to the $O6$-plane, an $F$-flux significant at the level of supergravity is needed. Condition~\C{massiveH}\ is somewhat similar to solving the heterotic Bianchi identity with a non-standard embedding so the right hand side of,
\be\label{hetbianchi}
d{\cal H}_{3}  = {\alpha' \over 4} \left\{ \tr R\wedge R  - \tr F\wedge F\right\},
\ee
is non-zero point-wise, though trivial in cohomology. However, the crucial difference is that a non-vanishing right hand side of~\C{hetbianchi}\ involves smooth sources of order $\alpha'$. The required non-closed ${\cal H}_3$ is of order $\alpha'$ and can be neglected at leading order in a large volume expansion. This is not the case for~\C{massiveH}.

Let us examine the dilaton equation~\C{massivedilaton}. The right hand side involves a localized source balanced against supergravity sources. If we set $\phi$ to a constant, there is no way to solve this equation. We must have a varying dilaton whose variation is significant at the level of the supergravity. However, an ${\cal N}=1$ supersymmetric $SU(3)$ structure solution requires a constant dilaton~\cite{Lust:2004ig, Lust:2009mb}, and a constant dilaton is assumed in~\cite{DeWolfe:2005uu}. A varying dilaton background might be possible for an $SU(3)\times SU(3)$ structure solution~\cite{Grana:2005sn, Lust:2009zb}. The proposed solutions of~\cite{DeWolfe:2005uu}\ therefore do not approximately solve the massive IIA supergravity equations of motion, which is a requirement for large volume, weakly-coupled backgrounds.

Indeed, one can go further and find more tension between the supersymmetry constraints and any solution that is a topologically trivial deformation of a Calabi-Yau metric with internal flux.  This should have been expected since Calabi-Yau metrics are not solutions of massive IIA in any approximation. Expanding around a large volume non-solution is bound to lead to trouble.

The current status of massive IIA compactifications can be summarized as follows: there are solutions of the general form described in~\cite{Lust:2009mb}\ with examples appearing in~\cite{Koerber:2008rx}. For this class of solutions, there is no large  separation of the $AdS_4$ scale from the Kaluza-Klein scale~\cite{Tsimpis:2012tu}, but there is still some chance that models with scale separation exist in supergravity, perhaps in the class studied in~\cite{Lust:2009zb}.

We do suspect that a mechanism similar to the one described in this work should exist for massive IIA, and such a mechanism might well allow very shallow $AdS_4$ solutions. However, unraveling that mechanism  will require some quantum understanding of massive IIA, its permitted sources, and the couplings those sources support.

\section*{Acknowledgements}

It is our pleasure to thank Bobby Acharya, Tom Banks, Katrin Becker, Francesco Benini, Greg Moore, Daniel Robbins, Michael Schulz, Mark Stern, Alessandro Tomasiello and Dimitrios Tsimpis for helpful discussions. S.~S. would like to thank the Isaac Newton Institute for Mathematical Sciences and the Simons  Center for Geometry and Physics for hospitality during the completion of this work.

\vskip 0.1in
J.~M. is supported in part by EPSRC Postdoctoral Fellowship EP/G051054/1. S.~S. is supported in part by
NSF Grant No.~PHY-0758029 and NSF Grant No.~0529954.

\newpage
\appendix
\section{T-duality Rules}\label{tduality}

For a string background specified by a metric $g$, $B$-field and dilaton $\phi$ with isometry in the $y$-direction, T-duality applied to the background gives a new background in terms of primed fields,
\bea
g'_{yy} = {1\over g_{yy}}, \qquad g'_{\mu y} = {B_{\mu y} \over g_{yy}}, \qquad g'_{\mu\nu} = g_{\mu\nu} - {g_{\mu y} g_{\nu y} - B_{\mu y} B_{\nu y} \over g_{yy}}, \\
B'_{\mu y} = {g_{\mu y} \over g_{yy}}, \qquad B'_{\mu\nu} = B_{\mu\nu} - {B_{\mu y} g_{\nu y} - g_{\mu y} B_{\nu y}\over g_{yy}}, \qquad \phi' = \phi - {1\over 2} {\rm log}( g_{yy}).
\eea
Any RR potentials in the original background transform as follows:
\bea
&& C^{(p)'}_{\mu_1 \cdots \mu_{p-1} y}  = C^{(p-1)}_{\mu_1 \cdots \mu_{p-1}} - (p-1) { C^{(p-1)}_{[\mu_1 \cdots \mu_{p-2} |y|} g_{\mu_{p-1}]y} \over g_{yy}}, \\
&& C^{(p)'}_{\mu_1 \cdots \mu_{p}} = C^{(p+1)}_{\mu_1 \cdots \mu_{p} y} + p C^{(p-1)}_{[\mu_1 \cdots \mu_{p-1}} B_{\mu_p]y} + p(p-1) {C^{(p-1)}_{[\mu_1 \cdots \mu_{p-2} |y} B_{\mu_{p-1}|y|} g_{\mu_p]y} \over g_{yy}}.
\eea

\newpage
%\bibliographystyle{amsunsrt-ensp}
%\bibliography{myrefs}

\ifx\undefined\bysame
\newcommand{\bysame}{\leavevmode\hbox to3em{\hrulefill}\,}
\fi

\end{document}